\documentclass[twocolumn,aps,prl,showpacs,amsmath,amssymb,superscriptaddress]{revtex4}

\usepackage{graphicx}
\usepackage{amsmath,amssymb}
\usepackage{textcomp}

\newcommand{\ket}[1]{\ensuremath{\left|  #1 \right\rangle}}
\newcommand{\var}[1]{\ensuremath{\left( \Delta #1 \right)^2}}

\begin{document}

\title{Collective State Measurement of Mesoscopic Ensembles with Single-Atom Resolution}

\author{Hao Zhang}
\affiliation{
Department of Physics, MIT-Harvard Center for Ultracold Atoms,
and Research Laboratory of Electronics, Massachusetts Institute of Technology,
Cambridge, Massachusetts 02139, USA}

\author{Robert McConnell}
\affiliation{
Department of Physics, MIT-Harvard Center for Ultracold Atoms,
and Research Laboratory of Electronics, Massachusetts Institute of Technology,
Cambridge, Massachusetts 02139, USA}

\author{Senka \'{C}uk}
\affiliation{
Department of Physics, MIT-Harvard Center for Ultracold Atoms,
and Research Laboratory of Electronics, Massachusetts Institute of Technology,
Cambridge, Massachusetts 02139, USA}
\affiliation{
Institute of Physics, University of Belgrade, Pregrevica 118, 11080 Belgrade, Serbia}

\author{Qian Lin}
\affiliation{
Department of Physics, MIT-Harvard Center for Ultracold Atoms,
and Research Laboratory of Electronics, Massachusetts Institute of Technology,
Cambridge, Massachusetts 02139, USA}

\author{Monika H. Schleier-Smith}
\affiliation{
Department of Physics, MIT-Harvard Center for Ultracold Atoms,
and Research Laboratory of Electronics, Massachusetts Institute of Technology,
Cambridge, Massachusetts 02139, USA}

\author{Ian D. Leroux}
\affiliation{
Department of Physics, MIT-Harvard Center for Ultracold Atoms,
and Research Laboratory of Electronics, Massachusetts Institute of Technology,
Cambridge, Massachusetts 02139, USA}

\author{Vladan Vuleti\'{c}}
\affiliation{
Department of Physics, MIT-Harvard Center for Ultracold Atoms,
and Research Laboratory of Electronics, Massachusetts Institute of Technology,
Cambridge, Massachusetts 02139, USA}

\date{\today}

\begin{abstract}
We demonstrate single-atom resolution, as well as detection sensitivity more than 20 dB below the quantum projection noise limit, for hyperfine-state-selective measurements on mesoscopic ensembles containing 100 or more atoms. The measurement detects the atom-induced shift of the resonance frequency of an optical cavity containing the ensemble. While spatially-varying coupling of atoms to the cavity prevents the direct observation of a quantized signal, the demonstrated measurement resolution provides the readout capability necessary for atomic interferometry substantially below the standard quantum limit, and down to the Heisenberg limit.
\end{abstract}

\maketitle

The rapidly progressing field of quantum metrology takes advantage of entangled ensembles of particles to improve measurement sensitivity beyond the standard quantum limit (SQL) arising from quantum projection noise for measurements on uncorrelated particles. Spin-squeezed states \cite{KitagawaUeda1993,WinelandSqueeze1992} improve the measurement signal-to-noise ratio by redistribution of quantum noise, while GHZ states \cite{GHZOriginal1989,GHZ1990, BollingerNOON1996} enhance the signal via faster-evolving collective phase. GHZ states enable measurement at the Heisenberg limit, where noise-to-signal ratio scales with atom number $N$ as $1/N$ \cite{BollingerNOON1996}.

In both cases, very-high-precision readout is necessary to realize metrological gain. The performance of an entangled interferometer is determined not
by the intrinsic fluctuations of the quantum system after detection
noise subtraction, but by the full observed noise including detection
noise  \cite{FernholzAtomicSqueeze2008,LerouxCavitySqueeze2010,GrossAtomInf2010, RiedelChipSqueeze2010,LouchetChauvet2010,Hamley2012}. Thus, the best observed spin squeezing of 6~dB \cite{LerouxCavitySqueeze2010} in a spin-$\frac{1}{2}$ system, and 8~dB of spin-nematic squeezing in a spin-1 system \cite{Hamley2012}, were both limited by detection. For GHZ states, read-out of the collective phase requires a measurement of the parity of the population difference between two atomic states \cite{BollingerNOON1996}. A state-selective measurement of atom number with single-atom resolution, which can be used to implement parity detection, therefore represents an important enabling technique for metrology beyond the SQL.

An optical cavity can be used both to collect photons in a single mode \cite{Mabuchi1996,SchlosserSubPoisson2001,McKeever2004b,Haase2006,Teper2006, FortierAtomLoading2007, TerracianoCavityBurst2009,Boozer2006, Gehr2010, Bochmann2010,SchleierSmithSqueezing2010}, and to generate entangled states via light-mediated atom-atom interactions \cite{MolmerSorensenBadCavity,AndreLukinPRA2002,LerouxCavitySqueeze2010}. With respect to atom detection, counting of up to 4 atoms \cite{Mabuchi1996, SchlosserSubPoisson2001, McKeever2004b, Haase2006, Teper2006, FortierAtomLoading2007, TerracianoCavityBurst2009} and high-fidelity readout of the hyperfine state of a single neutral atom \cite{Boozer2006, Gehr2010, Bochmann2010} have been achieved using cavity transmission measurements. Larger ensembles containing up to $N=70$ atoms have been measured with atom detection variance $\var{N}=6$ \cite{BrahmsCavityMeasure2011}. Spin-squeezed states of atoms in a cavity have also been prepared \cite{SchleierSmithSqueezing2010, LerouxCavitySqueeze2010, ChenSqueezing2011}, and have enabled an atomic clock operating with variance a factor of 3 below the standard quantum limit \cite{LerouxSqueezedClock2010}.

Single-atom resolution has also been achieved via fluorescence detection in free space. In optical lattices, the parity of site occupation has been measured for up to 5 atoms per lattice site without internal-state discrimination \cite{NelsonImagingArray2007, BakrQuantumMicro2009, ShersonSingleAtom2010}. For strongly trapped ions in a Paul trap, the individual states of up to 14 trapped ions have been detected via fluorescence collection \cite{Monz14Qubit2011, IslamQuantumIsing2011}. However, for a constant number of scattered photons per atom, the atom number resolution for fluorescence measurements deteriorates with increasing atom number as $\var{N} \propto N$, while for transmission (or cavity reflection) measurements, which are based on the collective forward scattering of light, the atom number resolution is (under ideal conditions) independent of atom number \cite{Teper2006}. Absorption measurements in free space are typically limited by technical and photon shot noise at a variance $\var{N} \gtrsim 50$ \cite{GrossAtomInf2010,RiedelChipSqueeze2010}. Measurements in free space have achieved variances surpassing the SQL by up to 9 dB in absorption \cite{GrossAtomInf2010} and up to $\sim$ 10 dB in fluorescence measurement \cite{Hamley2012}.

In this Letter, we demonstrate cavity-based high-fidelity state detection for mesoscopic ensembles. We achieve hyperfine-state-selective single-atom resolution for up to 100 atoms, and a measurement variance that is 21~dB below the projection noise limit already for a few hundred atoms. These represent improvements by more than an order of magnitude in atom number for single-atom resolution \cite{Mabuchi1996, SchlosserSubPoisson2001, McKeever2004b, Haase2006, Teper2006, FortierAtomLoading2007, TerracianoCavityBurst2009}, and by $\sim 10$~dB in measurement variance (relative to the SQL) over the best previous detection \cite{Hamley2012}. While spatially varying coupling of atoms to the probe light standing wave prevents direct observation of a quantized atom number signal, we demonstrate the ability to measure differences of one atom in our system and hence to perform the parity measurement that would detect a GHZ state in a uniformly coupled system \cite{Vrijsen2011}. When combined with entangled-state preparation by unitary cavity squeezing as proposed in \cite{QuantumEraser,MolmerSorensenBadCavity}, the demonstrated state readout will enable metrology substantially below the SQL, and at or near the Heisenberg limit.

\begin{figure}[tbph]
\centering
\includegraphics[width=.45 \textwidth]{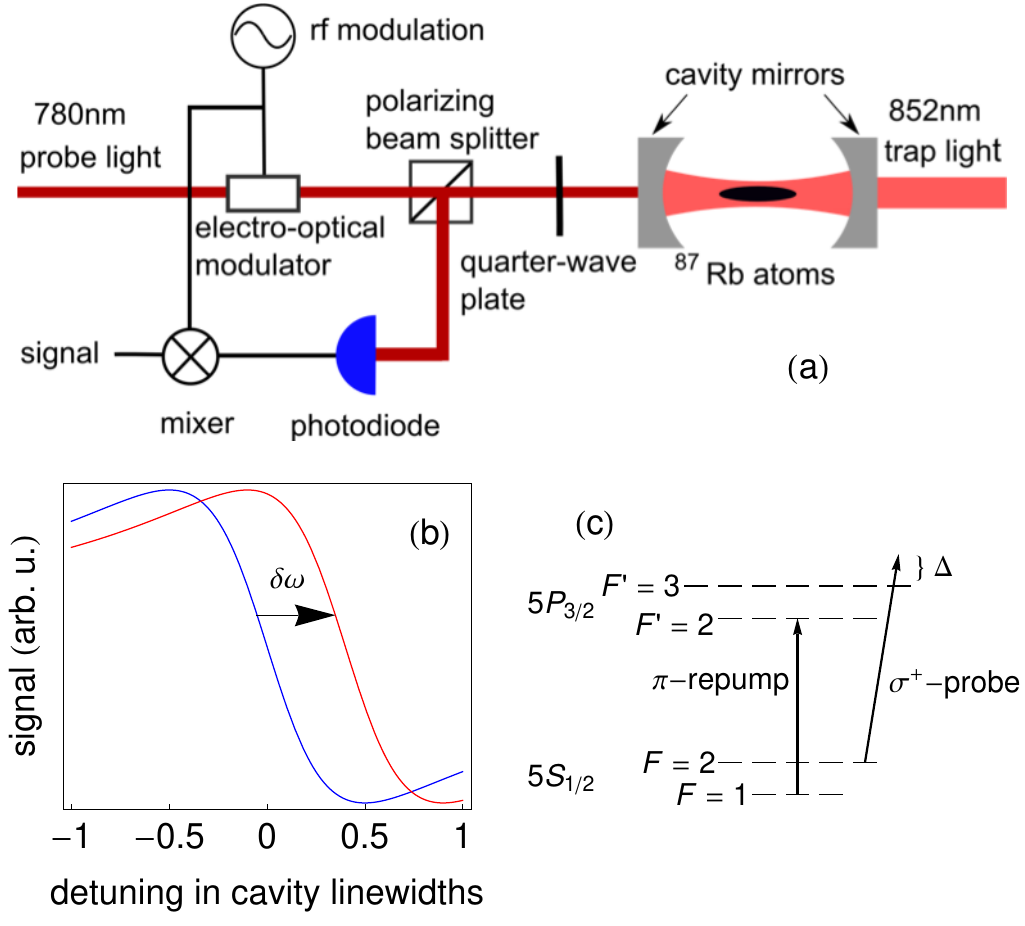}
\caption{Experimental setup. (a) Atoms are confined in a cavity by far-off-resonant 852 nm trap light, while a near-detuned probe beam is used to determine the cavity resonance frequency, which has been shifted by the atoms. (b) Pound-Drever-Hall signal produced by the heterodyne detection method (blue). The atoms' index of refraction shifts the cavity resonance by an amount proportional to the atom number $N$ (red). (c) Simplified level structure of $^{87}$Rb, showing the  $\ket{F=2} \rightarrow \ket{F'=3}$ probe light, and the $\ket{F = 1} \rightarrow \ket{F'=2}$ repumping light for total atom number measurements.}
\label{fig:Setup}
\end{figure}

We probe the atoms with near-resonant light of wavelength $2\pi/k=780$~nm inside a standing-wave optical cavity, while the atoms are trapped in a far-detuned intracavity standing wave of wavelength $2\pi/k_t=852$~nm. Thus an atom at a trap antinode at position $x$ is coupled with strength $g(x)=g_0 \cos(kx)$ to the probe field, where $2 g_0$ is the single-photon Rabi frequency at the probe antinode. When such an atom interacts with the cavity mode at large detuning $\Delta$ from the atomic resonance compared to the excited-state linewidth $\Gamma$, the cavity resonance is shifted by an amount $\omega_0 \cos^2(kx)$, where $\omega_0 = g_0^2/\Delta$ is the cavity shift by an atom at a probe antinode \cite{TanjiAdvancesCavity2011}. In the following, we specify atom number and noise measurements in units of maximally coupled atoms via the observed cavity shift $\delta \omega$ as $N= \delta \omega / \omega_0$. (The average actual atom number is twice as large.)

The cavity shift is measured using the Pound-Drever-Hall method \cite{PoundDreverHall1983}, which detects the phase of the light reflected from the cavity (Figure \ref{fig:Setup}). An electro-optical phase modulator operating at 127.5 MHz at a modulation index of $0.04$ produces sidebands on the probe light. The first-order red sideband is approximately resonant with the cavity, while the carrier serves as a phase reference. The reflected signal is heterodyned with the modulation source to produce a dispersive frequency-dependent signal in the vicinity of the cavity resonance. Compared to previous transmission measurements on the slope of the cavity resonance, the present measurement takes advantage of the twice stronger atom-cavity coupling on cavity resonance and the lower technical noise of a radiofrequency detection technique. We also operate at smaller detuning ($\Delta / (2 \pi) \leq 400$~MHz) than in Refs. \cite{SchleierSmithSqueezing2010, LerouxCavitySqueeze2010}, which results in a greater cavity shift per atom $\omega_0$, and reduces the impact of technical noise. Furthermore, by probing on the closed $|2, 2 \rangle \rightarrow |3', 3 \rangle$ transition, we limit Raman scattering to other states and increase the time over which we can measure.

The experimental setup is similar to the one previously used for spin squeezing and extensively characterized in Refs. \cite{SchleierSmithSqueezing2010, LerouxCavitySqueeze2010}. We confine 10 to 500 laser-cooled $^{87}$Rb atoms in a near-confocal cavity of free spectral range 5632(1) MHz and cavity linewidth $\kappa/(2 \pi) =1.01(3)$~MHz at the probe wavelength of 780 nm. The atoms are cooled in the trap of depth $U/ h = 18(3)$~MHz via polarization gradient cooling to a radial temperature $k_B T/h=1.0(1)$ MHz, confirmed via time-of-flight measurement, such that the radial rms cloud size of $\rho_{\mathrm{rms}} =7.0(7) \mathrm{\mu m}$ is much less than the probe light mode waist $w = 56.9(4) \mathrm{\mu m}$ at the atoms' position. The $\sigma^{+}$-polarized 780-nm probe beam with typical incident power of 2 nW enters the cavity and drives the $\ket{S_{1/2}, F=2, m_{F}=2} \rightarrow \ket{P_{3/2}, F'=3, m_{F}=3}$ transition with maximum single-photon Rabi frequency $2g_0/(2\pi)=1.12(4)$ MHz at the probe standing-wave antinode, as determined from first principles and accurately measured cavity parameters \cite{SchleierSmithSqueezing2010}. At the typical atom-cavity detuning $\Delta /(2 \pi)=250$~MHz, the cavity resonance is shifted $\omega_0 /(2\pi)=1.25(3)$ kHz per antinode atom. The probe light intracavity power is 12 $\mathrm{\mu W}$ and the depth of the potential generated by the probe is less than $h\times 3$ MHz, such that the probe light does not appreciably reduce the signal by pulling the atoms towards the nodes of the probe standing wave. (We also verify this experimentally by measuring the signal versus probe power.) Repumping light on the $\ket{F= 1} \rightarrow \ket{F^{\prime}=2}$ transition can be applied to optically pump the atoms into the $\left| F=2\right\rangle$ manifold. The probe light then pumps the atoms into the $\ket{F=2, m_{F}=2}$ state with respect to a 6.8 G magnetic field applied along the cavity axis.

The atoms' index of refraction can be thought of as arising from collective forward scattering of light by the ensemble, which makes the resolution in transmission or reflection measurements independent of atom number \cite{Teper2006}. In particular, for an ideal photon-shot-noise limited system, the detection variance is given by $\var{N}=(2 q \eta p)^{-1}$, where $p$ is the photon number scattered into free space per atom, $\eta = 4g_0^2/(\kappa \Gamma)=0.203(1)$ is the single-atom cooperativity (ratio of cavity to free-space scattering \cite{TanjiAdvancesCavity2011}) and $q = 0.2$ is the quantum efficiency of the detection (including detection path losses). In a typical measurement of duration $500 \, \mathrm{\mu}$s we scatter $p \approx 100$ photons per atom into free space. The accompanying recoil heating results in atom loss from the trap with a typical time constant of 30~ms.

\begin{figure}[htbp]
\centering
\includegraphics[width=.45 \textwidth]{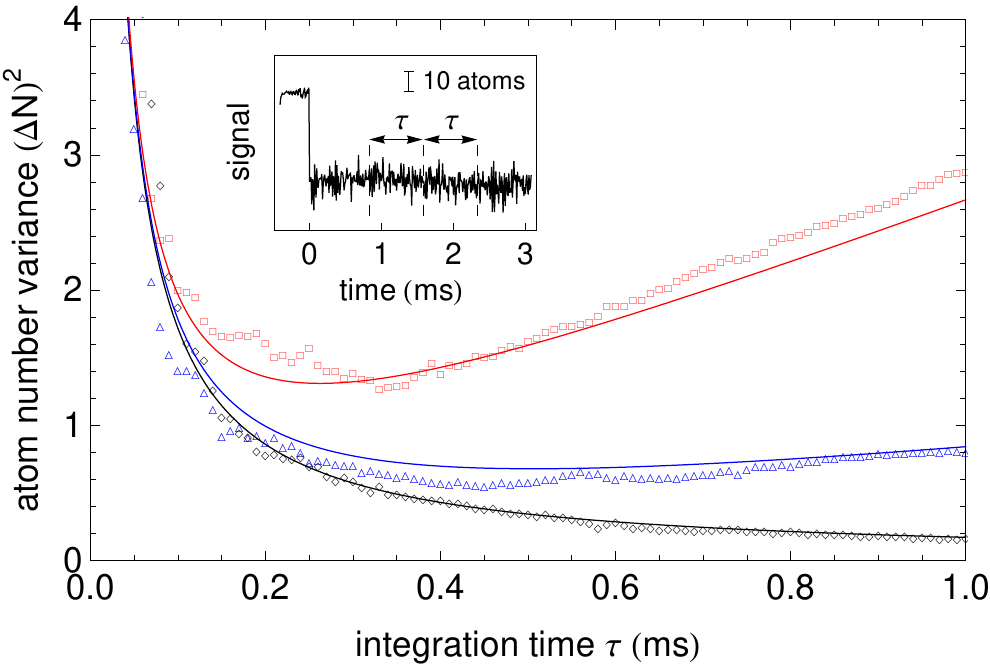}
\caption{Measured atom number variance as a function of integration time for $N=110$ atoms in $F = 2$ (red squares),  $N=30$ atoms in $F = 2$ (blue triangles), and an empty cavity with no atoms (black diamonds). The probe detuning is $\Delta /(2 \pi)=250$ MHz. The solid lines are fits according to Equation \ref{eqn:VarianceTime}. Inset: Typical signal trace. At $t = 0$ we introduce the probe laser to the cavity and observe a Pound-Drever-Hall signal. We integrate the signal over two periods of equal length $\tau$ and take the difference $N - N'$ to find the variance in atom number.}
\label{fig:IntegrationTime}
\end{figure}

We measure the atom number within the first 2 ms after the probe light enters the cavity. We average the signal for time $\tau$ to determine the shift $\delta \omega$ of the cavity resonance, which in turn determines the number of atoms $N=\delta \omega/\omega_0$. The difference in inferred atom number $N-N'$ between two adjacent detection bins allows us to determine the atom number variance at integration time $\tau$. The atom number resolution $\Delta N$ of our detector for a given $\tau$ is then given by $\var{N} =  \mathrm{Var}(N-N')/2$, where the variance is extracted from 100 repeated measurements. Figure \ref{fig:IntegrationTime} shows $\var{N}$ vs. integration time $\tau$ for  typical experimental parameters. The dependence is well described by the following model,
\begin{equation}
\var{N} =  c_1 \tau^{-1} + c_2 N \tau,
\label{eqn:VarianceTime}
\end{equation}
where $c_1$ and $c_2$ are constants that depend on the detuning $\Delta$ from the atomic transition, but not on atom number $N$. $c_1/\tau$ arises from photon shot noise or laser frequency noise \cite{SchleierSmithSqueezing2010}, while $c_2 N \tau$ represents the shot noise of the random atom loss. In principle there is also a fixed contribution to the variance due to technical noise, but we find this to be negligible. At each atom number and atom-cavity detuning $\Delta$ an optimal integration time $\tau$ can be found which minimizes Equation 1. The fitted $c_1$ coefficient reveals that our detection is a factor of three less sensitive than the photon shot noise limit, due to remaining frequency noise between laser and cavity. An effect not included in Equation \ref{eqn:VarianceTime} that becomes noticeable at larger atom number $N \gtrsim 50$ is the probe-laser-induced parametric heating of the atoms \cite{SchleierSmithCooling2011}, a collective optomechanical effect that gives rise to signal oscillations at twice the radial trapping frequency.

We investigate the dependence of the atom number resolution $\var{N}$ for both hyperfine-state selective and total atom number measurements. The probe directly detects only atoms in the hyperfine manifold $F=2$. In our state-selective measurement, probe polarization impurity leads to optical pumping of the atoms into the $F=1$ manifold, removing them from the measurement and contributing shot noise fluctuations. With probe light polarization purity better than 98\%, we observe at $\Delta/(2 \pi)=250$~MHz a time constant of 30~ms for decay to $\ket{F=1}$, corresponding to typically $6 \times 10^3$ scattered photons on the $\ket{2,2}\rightarrow \ket{3',3}$ transition. To measure total atom number, all atoms are kept in the state $\ket{2,2}$ by maintaining the repumping light during the measurement. In this case we do not expect shot noise from optical pumping to the $F = 1$ manifold.

\begin{figure}[tbph]
\centering
\includegraphics[width=.45 \textwidth]{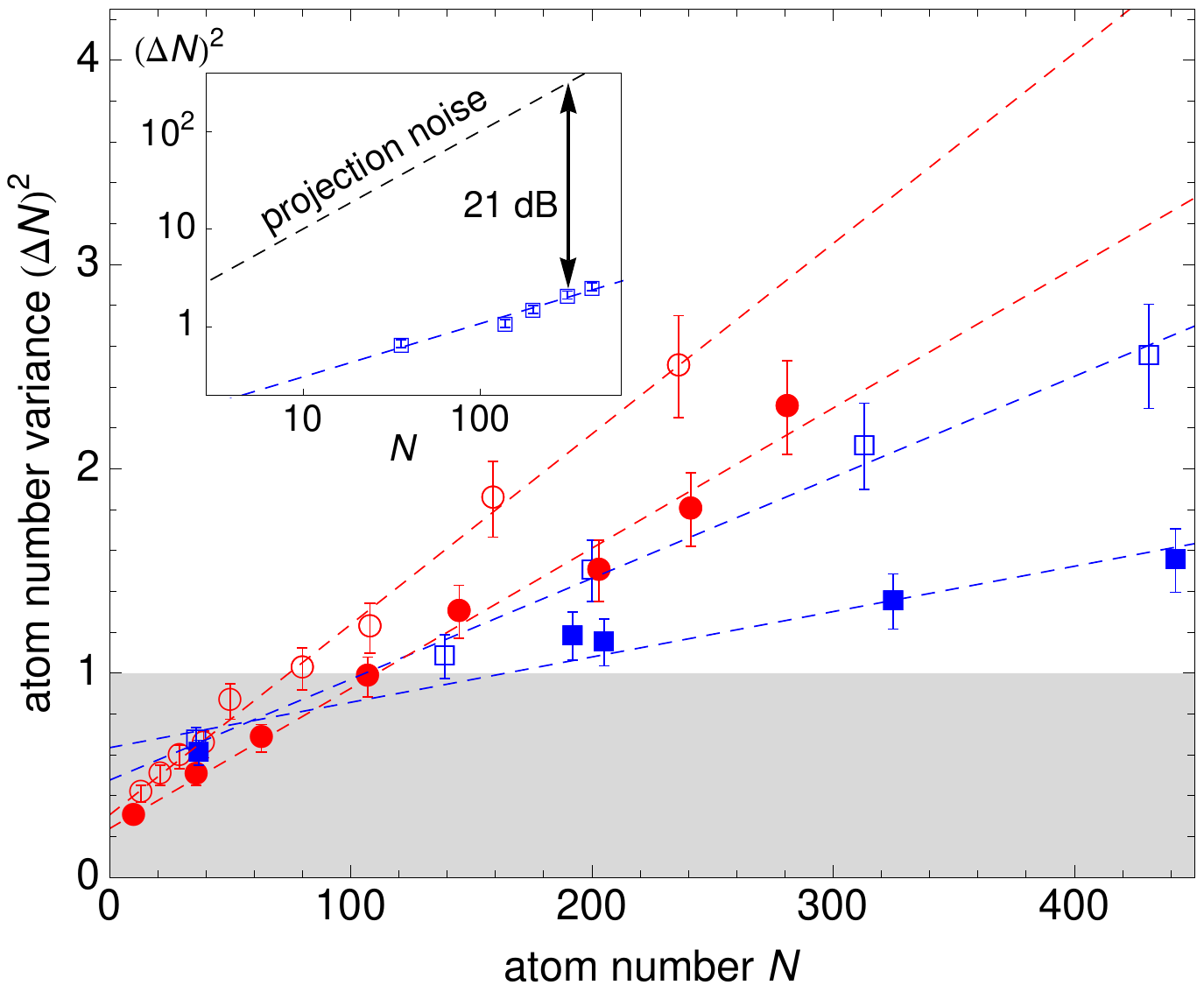}
\caption{Detection variance as a function of atom number for probe-atom detunings of $\Delta / (2 \pi) = 250$ MHz (red circles) and $\Delta / (2 \pi) = 375$~MHz (blue squares). Open (closed) symbols correspond to hyperfine-state-sensitive detection (total atom number detection). The shaded region indicates single-atom resolution, $(\Delta N)^2 <1$. Inset: Comparison of the measured detection variance to the projection noise, $(\Delta N)^2 = N$, that would be observed for uniformly-coupled atoms. At $\Delta / (2 \pi) = 375$~MHz, our detection is 20 dB below the projection noise already for 100 atoms.}
\label{fig:AtomCountingResult}
\end{figure}

Figure \ref{fig:AtomCountingResult} shows the variance $\var{N}$ for state-resolved detection and total atom number detection versus the number $N$ of atoms for two different light-atom detunings $\Delta$. (For each atom number, $\var{ N}$ is obtained from the minimum of a curve as displayed in Figure \ref{fig:IntegrationTime} that has itself been calculated from 100 repeated measurements.) In state-dependent detection, we obtain single-atom resolution $\var{N} < 1$ for up to 100 trapped atoms in the $\ket{2, 2}$ state. For total atom number detection, where shot noise from optical pumping to $F=1$ is absent, we retain single-atom resolution for $N \leq 150$. For smaller atom number, the resolution is better at smaller detuning $\Delta$, as we can scatter more photons and obtain more information before technical noise, predominantly due to slow drifts in our system which take place on the time scale of 1 ms, increases the signal variance. For larger atom number, optomechanical parametric oscillations \cite{SchleierSmithCooling2011} with period $\sim 200 \, \mu$s set a lower bound on the integration time, so that reducing the rate of heating and atom loss by increasing $\Delta$ improves the detection. Optomechanical heating is also responsible for the observed linear increase in detection variance  $\var{N} \propto N$ in Figure \ref{fig:AtomCountingResult} (which would otherwise scale as $\sqrt{N}$, as suggested by Equation \ref{eqn:VarianceTime}). For approximately 100 atoms, our state-selective resolution is already 20 dB below the quantum projection noise, $(\Delta N)^2 = N$, that sets the SQL. Nonuniform coupling of atoms to the probe beam and finite radial temperature combine to reduce the observed variance by 29\% compared to what would be observed in an ensemble of uniformly and maximally coupled atoms (see Supplementary Information), but in the latter case we would retain single-atom resolution for up to 70 atoms in state-selective detection, with variance 19 dB below the SQL at 100 atoms.

\begin{figure}[tbph]
\centering
\includegraphics[width=.45 \textwidth]{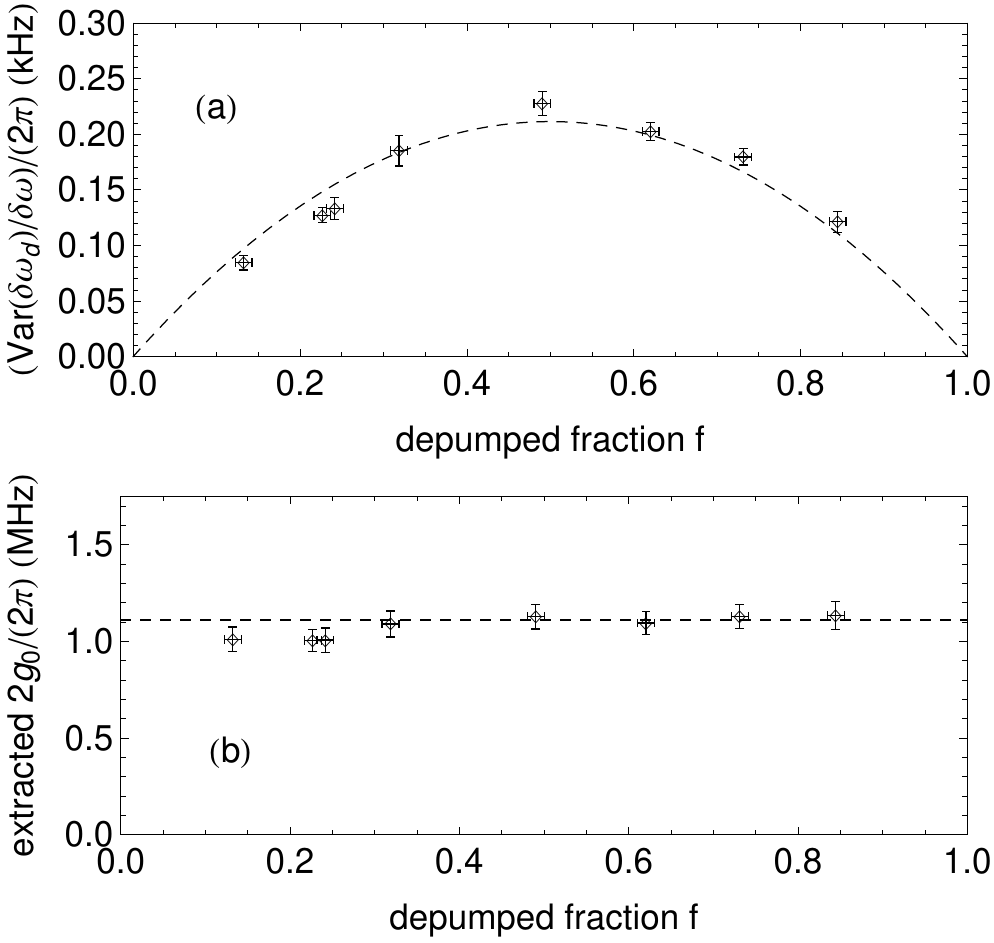}
\caption{Observation of the variance of a binomial distribution and verification of single-atom signal. (a) For an ensemble of $N=250$ atoms with cavity shift $\delta \omega$, a fraction $f$ is removed on average by optical pumping. The observed change in cavity shift, $\delta \omega_d$, is measured and the normalized variance $ V=\textrm{Var}(\delta \omega_d)/\delta \omega$ is plotted as a function of $f$. The dashed line is a model prediction with no free parameters of the form $V = \frac{3}{4} \alpha (g_0^2/\Delta) \,f \, (1-f)$, with $\alpha = 0.94$ (see text), and confirms the cavity shift $\omega_0=g_0^2/\Delta$ per atom. (b) The extracted value of the single-photon Rabi frequency $2 g_0$ from each point agrees with the value in our system calculated from first principles and independently measured cavity parameters (dashed line).}
\label{fig:DepumpVariance}
\end{figure}

In our present system, the nonuniform axial coupling to the probe beam prevents direct observation of a quantized atomic signal. The claim of single-atom resolution then critically relies upon correct calibration of the signal per atom $\omega_0=g_0^2/\Delta$. While this signal can be calculated from first principles using independently measured cavity parameters \cite{TanjiAdvancesCavity2011}, and our present setup has been extensively characterized previously in Ref. \cite{SchleierSmithSqueezing2010}, we present here a novel method that relies on the first direct observation, to our knowledge, of the binomial distribution in a system with fixed total atom number. We load typically $N = 250$ atoms, measure the initial atom-induced cavity shift $\delta \omega$, remove a fraction $f$ of the atoms to the state $F=1$ via optical pumping with $\pi$-polarized laser light tuned close to the $\ket{F = 2} \rightarrow \ket{F = 2'}$ transition, and measure the change in cavity shift $\delta \omega_d$. In Figure \ref{fig:DepumpVariance} we plot the measured normalized variance, $ V=\textrm{Var}(\delta \omega_d)/ \delta \omega$, as a function of the removed fraction $f$. The number of depumped atoms follows a binomial distribution, leading to the normalized variance of the cavity shift described by $V = \frac{3}{4} \alpha (g_0^2 / \Delta) \, f \, (1-f)$ (dashed line), where the factor of 3/4 accounts for nonuniform coupling of atoms to the probe beam and $\alpha = 0.94$ represents a small correction due to the measured radial temperature (see Supplementary Information). This model without any free parameters agrees very well with our data, as shown in Figure \ref{fig:DepumpVariance}(a). From each measured data point we can also directly extract the single-photon Rabi frequency at an antinode, $2g_0=4\sqrt{V \Delta/(3 \alpha f (1-f))}$, plotted in Figure \ref{fig:DepumpVariance}(b). A weighted average of extracted values gives $2 g_0/(2 \pi) = 1.08(2)$ MHz, in excellent agreement with the value 1.12(4)~MHz calculated from cavity parameters. This confirms the signal per atom, and, in combination with Figure \ref{fig:IntegrationTime}, the single-atom resolution of our measurement.

In conclusion, we have demonstrated the capability to detect differences of one atom in hyperfine-state occupation for ensembles of up to 100 atoms via measurement of the cavity resonance frequency. The demonstrated sensitivity enables the parity measurement that characterizes a GHZ state, with parity fringe visibility of about 30\% for 20-30 atoms and 50\% for a few atoms. Uniform atom-cavity coupling, required to observe a quantized atomic signal in the cavity system, can be achieved by using a trap wavelength that equals twice the probe wavelength \cite{Vrijsen2011}. The generation of GHZ states via atom-cavity interaction \cite{AndreLukinPRA2002, QuantumEraser} will require strong atom-cavity coupling $\eta>1$ to avoid decoherence by free-space scattering, and the readout resolution is likely to further improve in such a system.

This work was supported by the NSF, DARPA (QUASAR) and MURI through ARO, and the NSF Center for Ultracold Atoms. S.\ C. acknowledges support from the Ministry of Education and Science of the Republic of Serbia, through Grant No. III45016.

\end{document}